\documentclass[aip,reprint,english,a4paper] {revtex4-1}
\usepackage{ragged2e}
\usepackage{booktabs}
\usepackage{upgreek}
\usepackage{amsfonts}
\usepackage{amstext}
\usepackage{mathtools}
\usepackage{amssymb}
\usepackage{color}
\usepackage[normalem]{ulem}
\usepackage{amsmath} 
\usepackage{graphicx} 
\usepackage{bbold}

\AtBeginDocument{%
  \heavyrulewidth=.08em
  \lightrulewidth=.05em
  \cmidrulewidth=.03em
  \belowrulesep=.65ex
  \belowbottomsep=0pt
  \aboverulesep=.4ex
  \abovetopsep=0pt
  \cmidrulesep=\doublerulesep
  \cmidrulekern=.5em
  \defaultaddspace=.5em
}

\newcommand{\tabClusterCoords}   {S1}

\newcommand{\figRMSDorderpara}  {S1}
\newcommand{\figHelixUnfolding} {S2}
\newcommand{\figCaTimeEvol}     {S3}
\newcommand{\figClusterCoords}  {S4}
\newcommand{\figSingleTrajs}    {S5}

\begin{document}

\title{Nonequilibrium Modeling of the Elementary Step 
 in PDZ3 Allosteric Communication}
\author{Ahmed A.\ A.\ I.\ Ali}
\altaffiliation{A.\ Ali and A.\ Gulzar contributed equally
  to this work.} 
\author{Adnan Gulzar}
\altaffiliation{A.\ Ali  and A.\ Gulzar contributed
  equally to this work.} 
\author{Steffen Wolf}
\author{Gerhard Stock}
\email{stock@physik.uni-freiburg.de}
\affiliation{Biomolecular Dynamics, Institute of Physics,
   University of Freiburg, 79104 Freiburg, Germany}
\date{\today}

\begin{abstract}

{\bf ABSTRACT:} While allostery is of paramount importance for protein
signaling and regulation, the underlying dynamical process of allosteric
communication is not well understood. PDZ3 domain represents a prime
example of an allosteric single-domain protein, as it features a
well-established long-range coupling between the C-terminal
$\alpha_3$-helix and ligand binding. In an intriguing experiment,
Hamm and coworkers
employed photoswitching of the $\alpha_3$-helix to initiate a
conformational change of PDZ3 that propagates from the C-terminus to
the bound ligand within 200 ns.
Performing extensive nonequilibrium molecular dynamics simulations,
the modeling of the experiment reproduces the measured timescales and
reveals a detailed picture of the allosteric communication in PDZ3. In
particular, a correlation analysis identifies a network of contacts
connecting the $\alpha_3$-helix and the core of the protein, which
move in a concerted manner. Representing a one-step process and
involving direct $\alpha_3$-ligand contacts, this cooperative
transition is considered as elementary step in the propagation of
conformational change.

\end{abstract}
\maketitle

{\centering \includegraphics[width=0.45\textwidth]{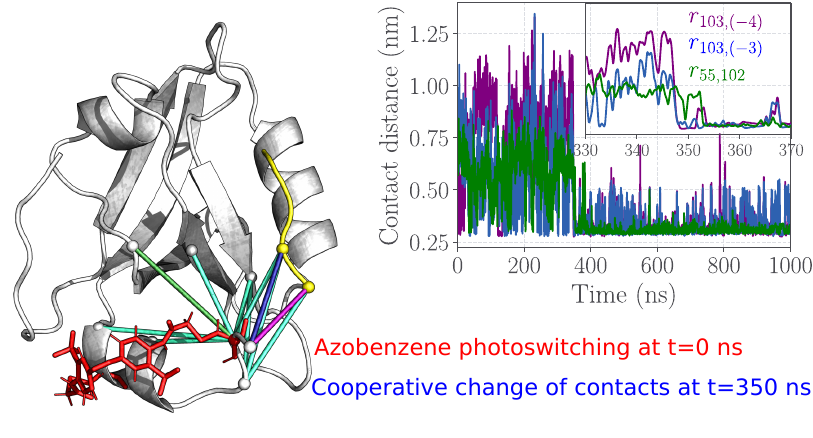}}
\vspace*{-6mm}


%
%
\section*{Introduction}
\vspace*{-4mm}

\begin{figure}[ht!]
\centering
\includegraphics[width=0.49\textwidth]{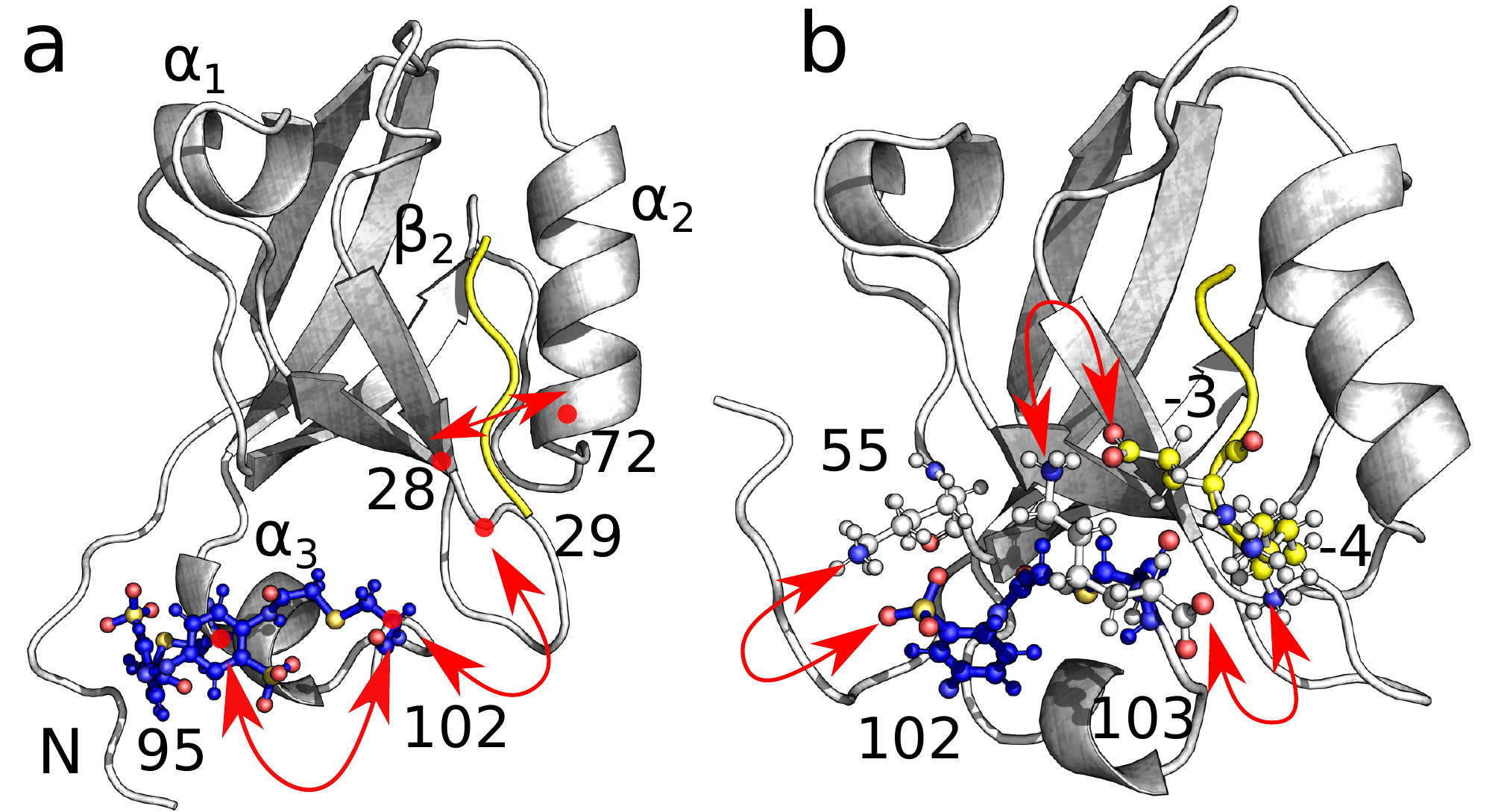}
\caption{
Photoswitchable PDZ3 domain studied by
  Bozovic et al.\cite{Bozovic21} Upon {\em cis}-to-{\em trans}
  isomerization, the azobenzene photoswitch (blue) unfolds the
  $\alpha_3$-helix of the system and initiates an allosteric
  transition that changes the binding affinity of the ligand (yellow).
  Panel (a) shows the {\em cis} state of the system and indicates main
  secondary structural elements as well as $C_{\alpha}$-distances
  $d_{95,102}$, $d_{28,72}$ and $d_{29,102}$ discussed in the
  text. Panel (b) shows the {\em trans} state with its three
  stabilizing salt bridges Azo102-Lys55. Lys103-Lys(-4) and
  Lys103-Glu(-3). Residues are numbered from 1 to 103 for the protein
  and from -4 to 0 for the ligand.}
\label{fig:pdz}
\end{figure}

Allostery is one of the most important mechanisms of biomolecular
regulation and signal transduction. \cite{Wodak19,Gunasekaran04,
  Bahar07,Cui08,Changeux12, McLeish13, Motlagh14,Tsai14,Thirumalai19}
While commonly this term is meant to describe the communication
between distal domains of large macromolecules, it has been suggested
that even relatively small single-domain proteins exhibit allosteric
properties. \cite{Gunasekaran04} PDZ domains, for example, are
well-established and structurally conserved protein interaction
modules involved in the regulation of multiple receptor-coupled signal
transduction processes, but at the same time have also been studied
extensively as isolated model systems of allosteric
communication. \cite{Fuentes04,Petit09, Lee10,Ye13} They share a
common fold, which consists of two $\alpha$-helices and six
$\beta$-strands, with the second $\alpha$-helix and the second
$\beta$-strand forming the canonical ligand binding groove (Fig.\
\ref{fig:pdz}), and generally bind the C-terminus of their targets.

Notably, the NMR study of Petit et al. \cite{Petit09} showed that the
removal of a short $\alpha$-helix at the C-terminal of PDZ3 reduces
ligand affinity by a factor of 21, thus revealing an allosteric
communication between ligand binding and C-terminal dynamics.  In this
way, PDZ3 can be regarded as one of the smallest allosteric
proteins, in the sense that both the active site (the
$\alpha_3$-helix) and the allosteric site (the binding pocket) are
clearly defined. Notwithstanding numerous experimental and
computational studies of PDZ domains (see Refs.\
\onlinecite{Fuentes04,Petit09, Lee10,Ye13, Lockless99,
  Ota05,Kong09,Gerek11,Ishikura15, Kumawat17, Stock18,Faure22} for a
rather incomplete selection), however, it is still not well understood
how exactly a local perturbation propagates through the protein in
space and time to a distal site.
This is due to the typically small local structural changes of an
allosteric signal which are challenging to observe in experiments,
\cite{Brueschweiler09} and also because of the timescale limitations
of molecular dynamics (MD)
simulations.\cite{Chen07,Vesper13,Pontiggia15,Zheng18} 

\vspace*{-4mm} To facilitate a real-time study of the underlying
allosteric transition in PDZ domains, Hamm and coworkers implemented
an azobenzene photoswitch that impulsively perturb the protein (e.g.,
by causing the unbinding of the ligand), and monitored the subsequent
structural evolution of the system via time-resolved vibrational
spectroscopy. \cite{Buchli13,Bozovic20,Bozovic20a,Bozovic21,Bozovic22}
As a main result, they found for PDZ2 that the conformational
rearrangement of the photoswitchable protein occurs on multiple
timescales from pico- to microseconds in a highly nonexponential
manner. \cite{Stock18} Accompanying MD studies of the nonequilibrium
dynamics reproduced these findings and revealed a quite complex
structural reorganization of the system. \cite{Buchenberg14,
  Buchenberg17} Most recently, Bozovic et al.\cite{Bozovic21} employed
photoswitching of the $\alpha_3$-helix to initiate a conformational
change of PDZ3 that propagates from the $\alpha_3$-helix to the
binding pocket of the ligand. They found a timescale of 4 to 6\,ns for
the enforced unfolding of the $\alpha_3$-helix, as well as 200\,ns for
the time it takes for the allosteric signal to reach the ligand.

To aid the interpretation of these experiments and to unravel the
atomistic mechanism of the putative allosteric transition in PDZ3, in
this work we perform extensive all-atom explicit-solvent MD
simulations of photoswitchable PDZ3. Aiming to directly model the
experiment of Bozovic et al.,\cite{Bozovic21} we employ a
potential-energy surface switching method\cite{Nguyen06} to collect
$100 \times 1\, \mu$s-long nonequilibrium trajectories of the
photoinduced conformational change of PDZ3. In excellent agreement
with experiment, we find that the initial unfolding of the
$\alpha_3$-helix occurs on a 5\,ns timescale and causes a population
shift of protein-attached and protein-separated conformations of the
$\alpha_3$-helix, which in turn triggers a long-range conformational
transition involving the ligand on a timescale of about 300\,ns.

%
%
\vspace*{-4mm}
\section*{Results}
\vspace*{-4mm}
\subsection*{Similarity of photoswitchable and wild-type PDZ3}
\vspace*{-4mm}

In the experiment of Bozovic et al.,\cite{Bozovic21} the azobenzene
photoswitch was attached to the side-chains of residues 95 and 102 of
the $\alpha_3$-helix. In this way, the end-to-end distance of
azobenzene in its stretched {\em trans} configuration accommodates a
stable $\alpha_3$-helix, while the twisted {\em cis} configuration
destabilizes the helix. To verify that the photoswitchable PDZ3
represents a suitable model for the wild-type protein, we performed
$8\times 10\,\mu$s-long equilibrium MD simulations of the {\em cis}
and the {\em trans} systems, based on a crystal
structure\cite{Doyle96} (PDB entry 1TP5) and using the GROMACS v2020
software package\cite{GROMACS20} and the Amber99SB*ILDN force
field,\cite{Hornak06, Best09, LindorffLarsen10} see SI
Methods. Calculating the time-dependent root mean square displacement
(RMSD) matrix of the {\em cis} and the wild-type trajectories, Fig.\
\figRMSDorderpara\ shows that the C$_\alpha$-RMSD of the full protein
never exceeds 2.65\,\AA\ throughout the $1\,\mu$s time evolution
shown. With an average of $1.45\,$\AA, the RMSD can be regarded as a
sufficient similarity between photoswitchable and wild-type protein.
As an further validation of the simulation model, we calculated the
side-chain methyl groups fluctuations\cite{Hoffmann20} as measured by
the NMR order parameters $S^2$ of Petit et al.,\cite{Petit09} see SI
Methods. Comparing $S^2$ values obtained from the simulation of {\em
  cis} PDZ3 and the wild-type protein as well as from
experiment and wild-type simulation, we obtain good overall
agreement (Fig.\ \figRMSDorderpara). 

%
%
\vspace*{-4mm}
\subsection*{Structural changes of PDZ3 upon photoswitching}
\vspace*{-4mm}

\begin{figure}[h!]
\centering
{\includegraphics[width=0.35\textwidth]{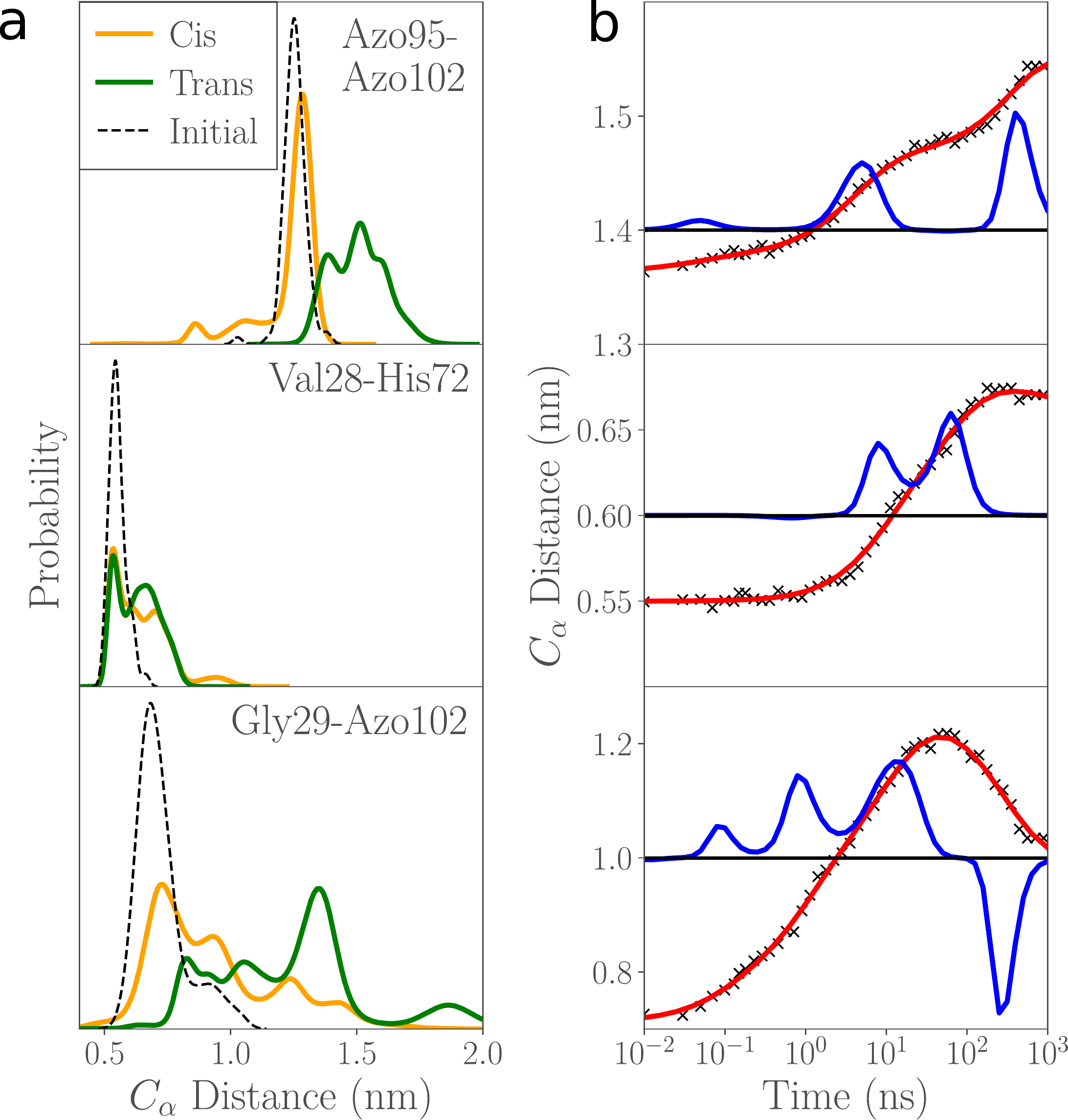}}
\caption{
  Distribution (a) and time evolution (b) of three
  C$_{\alpha}$-distances that describe the structural rearrangement of
  PDZ3: $d_{95,102}$ indicating the length of $\alpha_{3}$-helix,
  $d_{28,72}$ representing the width of the binding pocket, and
  $d_{29,102}$ accounting for the distance between $\rm{\alpha_{3}}$
  and the core of the protein. Distributions are obtained for the {\em
    cis} and {\em trans} equilibrium states, as well as for the
  initial state of the nonequilibrium simulations. The time traces in
  red are fits of the MD data (black points) from the timescale
  analysis [Eq.\ (\ref{EqMultiexp})], and the blue lines are the
  resulting timescale spectra.}
\label{fig:Ca}
\end{figure}

To get a first impression of the photoinduced conformational
transition of PDZ3, we follow Buchenberg et al.\cite{Buchenberg17} and
consider selected interresidue C$_{\alpha}$-distances that account for
various aspects of the structural rearrangement. As a first example,
Fig.\ \ref{fig:Ca} shows the distance $d_{95,102}$ between residues
Azo95 and Azo102 bridged by the photoswitch, which reports on the
length of the $\alpha_{3}$-helix (Fig.\ \ref{fig:pdz}a). The
distribution of $d_{95,102}$ reveals a shift of about 3\,\AA\ upon
switching from the {\em cis} to the {\em trans} equilibrium state,
which clearly reflects the photoinduced stretching and partial
unfolding of $\alpha_{3}$.

To study the time evolution of the system following {\em cis} to
{\em trans} photoswitching, we sampled 100 statistically
independent structures from the {\em cis} equilibrium simulations,
performed at time $t\!=\!0$ a potential-energy surface switching
method,\cite{Nguyen06} and calculated $100 \!\times \!1\, \mu$s-long
nonequilibrium trajectories. By performing an ensemble
average over these trajectories, we obtain the time-dependent mean
value $\langle d (t) \rangle$ of some observable $d$. Using a
log-scale representation of the time axis ranging from 10\,ps to
1\,$\mu$s, Fig.\ \ref{fig:Ca}b shows the photoinduced evolution of
$\langle d_{95,102} (t) \rangle$, which is found to increase on
several timescales. To facilitate the interpretation, we perform a
timescale analysis using a maximum entropy
method\cite{Lorenz-Fonfria06}
\begin{equation} \label{EqMultiexp}
\langle d (t) \rangle = \sum_k a_k e^{-t/\tau_{k}} ,
\end{equation}
where the amplitudes $a_k$ of the multiexponential fit yield the
timescale spectrum $a(\tau_{k})$ of the evolution.
We used 10 equally
distributed timescales $\tau_{k}$ per decade and a regularization
parameter $\lambda=3\times 10^{-4}$. The analysis of
$\langle d_{95,102} (t) \rangle$ reveals that the sub-ps
photoisomerization of azobenzene\cite{Naegele97} causes an elongation
of the $\alpha_{3}$-helix on timescales of $\sim 5$ and $400\,$ns. A
structural analysis (Fig.\ \figHelixUnfolding) shows that the
short timescale reflects the initial strtching of $\alpha_{3}$, which
is in perfect agreement with the experimental result (4 to 6\,ns) of
Bozovic et al.\cite{Bozovic21} The long timescale accounts for the
structural relaxation of the $\alpha_{3}$-helix due to the subsequent
rearrangement of PDZ3 to be discussed below.

As an indicator of how the initial stretching of the
$\alpha_{3}$-helix affects the environment of the distal ligand of
PDZ3, Fig.\ \ref{fig:Ca} shows the distance between Val28 in the
$\beta_2$-strand and His72 in the $\alpha_2$-helix, which accounts for
the width of the binding pocket\cite{Buchenberg17} (Fig.\
\ref{fig:pdz}a). While the distributions of $d_{28,72}$ in {\em cis}
and {\em trans} equilibrium states are rather similar, we find a clear
increase of $\langle d_{28,72} (t) \rangle$ on timescales of $\sim 8$
and $60\,$ns. This supposed discrepancy between distributions and time
evolution reflects the fact that the nonequilibrium simulations were
started close to the initial structure (dotted lines in Fig.\
\ref{fig:Ca}), and therefore exhibit a short-time relaxation to the
appropriate conformational distribution.

Finally, we consider the $C_{\alpha}$ distance between Azo102 and
Gly29 in the $\beta_2$-$\beta_3$ loop as an observable that reports on
the attachment of the $\alpha_{3}$-helix to the core of the protein
(Fig.\ \ref{fig:pdz}a). Since the alignment and the separation of
$\alpha_{3}$ to and from the protein core is believed to represent the
origin of the change in ligand binding
affinity,\cite{Petit09,Bozovic20a} $d_{29,102}$ can be considered as a
simple descriptor of the allosteric transition. Figure \ref{fig:Ca}a
shows that the distribution of $d_{29,102}$ is rather broad and
exhibits a significant shift ($\sim 0.7\,$nm) of its maximum upon {\em
  cis}-to-{\em trans} switching. When we roughly associate distances
$d_{29,102} \lesssim $1\,nm with structures where the
$\alpha_{3}$-helix is attached to the core, and larger distances with
an unattached $\alpha_{3}$-helix, $d_{29,102}$ suggests a decrease of
68 to 32\% of attached structures when we change from {\em cis} to
{\em trans}.
Rather than a simple one-to-one relation of {\em cis} and {\em
  trans} states with an attached and separated $\alpha_3$-helix,
respectively, we thus find a coexistence of both structures in both states,
which undergo a population shift upon {\em cis} to {\em trans} switching.

In line with its large structural heterogeneity, the time evolution of
the $\alpha_{3}$-core distance is found to be quite complex.
$\langle d_{29,102} (t) \rangle$ first increases on several timescales
($\sim 0.1$, 1 and 10\,ns), before it starts to decrease on a
$\sim \,250\,$ns timescale. While the fast timescales again reflect
the short-time relaxation of the initially prepared state, the long
timescale seems to indicate the subsequent allosteric transition of
PDZ3.
To summarize, we have identified a timescale of 5\,ns associated with
the initial stretching of the $\alpha_{3}$-helix, an intermediate
timescale (tens of ns) that reflects the short-time relaxation of the
initially prepared structural distribution, and a slow timescale
($\sim 300\,$ns) whose structural origin is to be explained next.

%
%
\vspace*{-4mm}
\subsection*{Correlation analysis of interresidue contacts}
\vspace*{-4mm}

To identify internal coordinates that change
significantly during the conformational transition, we follow recent
work\cite{Ernst17,Post22a} and focus on interresidue contact
distances. This is for two reasons. First, although contact distances
report only on near-order interactions, long-distance changes of the
structure result as a consequence and are therefore included as
well. Moreover, while side-chain dihedral angles may also mediate
allosteric couplings,\cite{Bowman12,Post22a} analyses including them
showed that they are not of relevance for PDZ3. Assuming that a contact is
formed if the distance $r_{ij}$ between the closest non-hydrogen atoms
of residues $i$ and $j$ is shorter than 4.5\,\AA,\cite{Ernst15}
we identified 403 interresidue contacts (see SI Methods). To
discriminate collective motions underlying functional dynamics from
uncorrelated motion, we calculated (the modulus of) the linear
correlation matrix of these coordinates and rearranged this matrix in
an approximately block-diagonal form. Following Diez et al.,
\cite{Diez22} this is achieved via a community detection technique
called Leiden clustering,\cite{Traag19} employing the Python package
MoSAIC\cite{Diez22} and a Leiden resolution parameter $\gamma = 0.5$.

\begin{figure}[ht!]
  \centering
  {\includegraphics[width=0.44\textwidth]{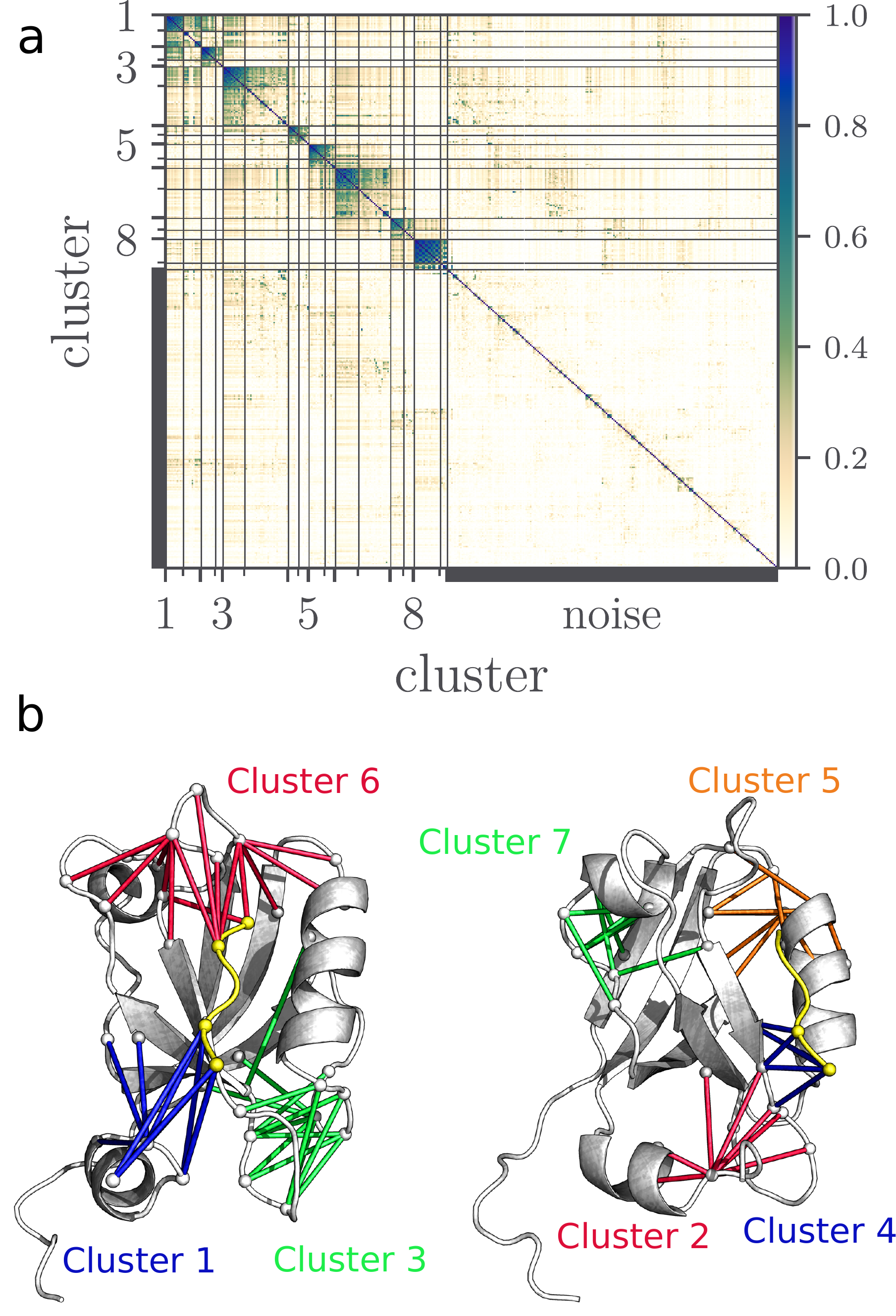}}
  \caption{
    (a) Block-diagonal correlation matrix obtained from Leiden
    clustering of 403 contact distances of PDZ3. In this way, we
    discriminate collective motions underlying functional dynamics
    (represented by the main clusters) from uncorrelated motion or
    noise (shown in the lower square). (b) Illustration of the contact
    distances (colored lines) associated with the main clusters.}
  \label{fig:Leiden}
\end{figure}

Figure~\ref{fig:Leiden}a shows the resulting block-diagonal
correlation matrix $\{\sigma_{ij}\}$, which reveals eight main blocks
or clusters. Within such a cluster, the coordinates are highly
correlated (i.e., on average $\sigma_{ij} \ge \gamma$), while the
correlation between different clusters is low (i.e.,
$\sigma_{ij} < \gamma$). On the other hand, we find that $\sim$56\,\%
of all coordinates (shown in the lower right square) correlate only
weakly with few other coordinates. (This is, e.g., the case for stable
contacts and contacts on the protein surface that form and break
frequently.)  These contact distances are therefore classified as
noise and can be omitted in the further analysis. Finally, there are
coordinates (or mini-clusters with $\le 3$ coordinates) that are not a
member of a main cluster but still exhibit moderate correlation with
some of the main clusters; they are shown in between the main
clusters.

To illustrate the coordinates contained in the main clusters, Fig.\
\ref{fig:Leiden}b shows the corresponding contact distances inserted
into the structure of PDZ3. (See Table \tabClusterCoords\ for a list
of the coordinates of all clusters.) Most interestingly, we find that
cluster\,1 contains contacts between the $\alpha_3$-helix and the
ligand as well as the core of the protein.  Causing the (un)alignment
of $\alpha_{3}$ to the protein core, which in turn induces the change
in ligand binding,\cite{Petit09,Bozovic20a} these coordinates are the
key to describe the allosteric transition in PDZ3. The remaining
clusters, on the other hand, are found to describe interactions in
different sections of PDZ3, whose motions are mostly uncorrelated to
the coordinates of cluster\,1.  For example, cluster\,2 connects
residue Phe100 of $\alpha_3$ to the $\beta_2$-$\beta_3$ loop, and
cluster\,3 and cluster\,6 accounts for motions within the
$\beta_2$-$\beta_3$ and $\beta_1$-$\beta_2$ loops, respectively.


\begin{figure}[h!]
  \centering
  {\includegraphics[width=0.4\textwidth]{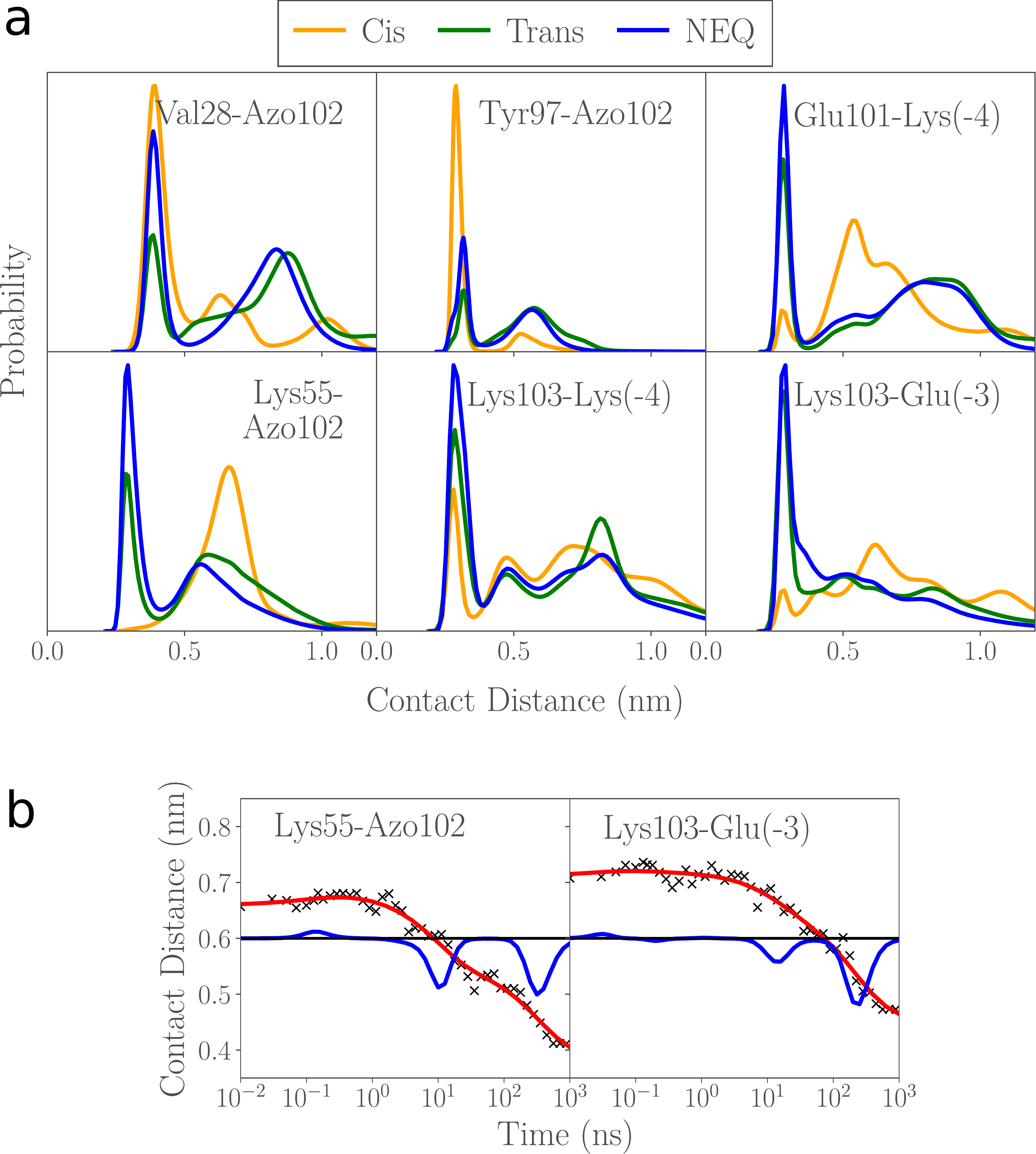}}
  \caption{
    (a) Distribution of various contact distances $r_{ij}$ that
    mediate the allosteric transition in PDZ3, obtained for the {\em
      cis} and {\em trans} equilibrium states, as well as during the
    nonequilibrium simulations. (b) Time evolution of the salt bridge
    distances $r_{55,102}$ and $r_{103,(-3)}$. Time traces in red are
    fits of the MD data (black points) from the timescale analysis
    [Eq.\ (\ref{EqMultiexp})], blue lines are the resulting timescale
    spectra.}
\label{fig:contacts}
\end{figure}

Hence the Leiden correlation analysis leaves us with 11 (instead of
initially 403) coordinates comprised in cluster\,1, which form a
network of highly correlated interresidue contacts (Fig.\
\ref{fig:Leiden}b). The network includes 6 hydrophobic contacts, 2
hydrogen bonds, and 3 salt bridges, which connect the $\alpha_3$-helix
(mostly residues Azo102 and Lys103) to ligand residues Glu(-3) and
Lys(-4) and various protein residues. Selecting six representative
contacts (see Fig.\ \figClusterCoords\ for the other cases), Fig.\
\ref{fig:contacts}a shows the corresponding distance distributions in
the {\em cis} and {\em trans} equilibrium states.
For all contacts we find a coexistence between
contact-formed states (with a well-defined peak at distances
$\lesssim 4.5\,$\AA) and contact-broken states (with a broad
distribution of longer distances). We note that in {\em cis} the
$\alpha_3$-helix is mostly stabilized by nonpolar contacts with the protein
(e.g., Val28 and Tyr97 connect to Azo102), while in {\em trans} polar
contacts with the ligand dominate. The latter include two salt bridges
between C-terminal Lys103 and the ligand residues Lys(-4) and Glu(-3),
as well as a hydrogen bond between the backbone of Glu101 and the
side-chain of Lys(-4). Moreover, we find the salt
bridge  Azo102-Lys55 between $\alpha_3$ and the protein core, which
exists almost exclusively in {\em trans}. That is, during the
allosteric transition, (at least) four new contacts are formed,
including three salt bridges and one hydrogen bond.

As representative examples of contacts that are formed during the
photoinduced transition, Fig.\ \ref{fig:contacts}b shows the time
evolution of the salt bridges Lys55-Azo102 and Lys103-Glu(-3). In both
cases, the formation occurs on two timescales of about 10\,ns and
300\,ns. Similar results are also obtained for the other contact
distances of cluster 2 (Fig.\ \figClusterCoords).  While the 10\,ns timescale
reflects the above discussed short-time relaxation of the initially
prepared structural distribution, the long timescale (already found in
Fig.\ \ref{fig:Ca}b) can now unambiguously be associated with the
allosteric transition.
Assuming that the experimentally measured infrared response of the
ligand is related to changes of protein-ligand contacts, the 300\,ns
timescale compares well to experimental ligand response time of
200\,ns obtained by Bozovic et al.\cite{Bozovic21}

\vspace*{-10mm}
%
%
\subsection*{Cooperative mechanism of the long-range communication}
\vspace*{-4mm}

\begin{figure}[h!]
  \centering
  {\includegraphics[width=0.35\textwidth]{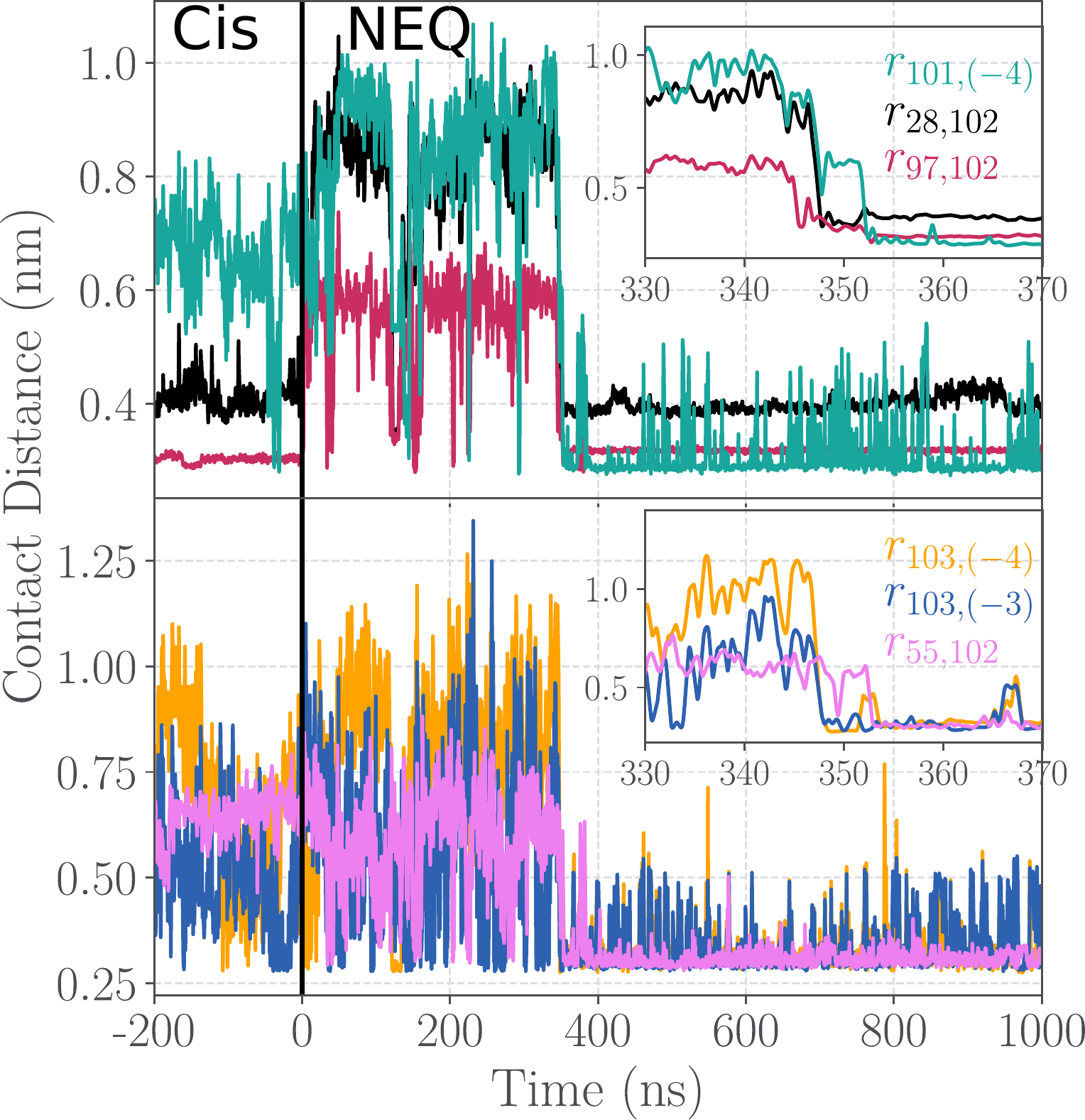}}
  \caption{
    Time evolution of selected contact distances along a {\em single}
    trajectory of PDZ3. For times -200\,ns $\le t \le 0$, the system
    is in the {\em cis} equilibrium state, and undergoes a
    nonequilibrium evolution following {\em cis}-to-{\em trans}
    photoswitching at $t=0$. For smoothing, a Gaussian filter with
    width $\sigma = 0.2\,$ns was used.}
\label{fig:SingleTraj}
\end{figure}

The above correlation analysis indicates that the allosteric
transition in PDZ3 is mediated by a network of 11 highly correlated
interresidue contacts (Fig.\ \ref{fig:Leiden}b). This raises the
question if the transition proceeds sequentially (i.e., involving
intermediate states) or rather in a concerted manner. To address this
point, Fig.\ \ref{fig:SingleTraj} shows the time evolution of the
above discussed contact distances along a {\em single} nonequilibrium
trajectory (i.e., without performing an ensemble average). Starting in
the {\em cis} equilibrium state for times -200\,ns $\le\! t\! \le 0$,
we find that the two {\em cis}-stabilizing contacts are formed (i.e.,
distances $r_{28,102}$ and $r_{97,102}$ fluctuate with small amplitude
around 3\,\AA), while the {\em trans}-stabilizing contacts are broken
such that the corresponding distances $r_{55,102}$, $r_{101,(-4)}$,
$r_{103,(-3)}$ and $r_{103,(-4)}$ fluctuate with large amplitudes.
Upon {\em cis}-to-{\em trans} photoswitching at $t\!=\!0$, contacts
28-102 and 97-102 are broken, too, within tens of nanosecond. As a
consequence, all distances fluctuate wildly, reflecting the
large-amplitude motion of the detached $\alpha_3$-helix. At
$t\sim 348\,$ns this motion localizes abruptly, when all six contacts
are formed within 5--7~ns, i.e., virtually simultaneously on the
transition timescale of hundreds of nanoseconds. While contact
distances $r_{102,(-4)}$, $r_{103,(-3)}$ still exhibit residual
fluctuations, the other contacts are tightly closed.

Inspecting the other nonequilibrium trajectories, Fig.\
\figSingleTrajs\ shows that the overall picture is qualitatively
similar to the above example. Of the $100 \times 1\,\mu$s-long
trajectories, 44 remain in the large-amplitude state, while the other
56 undergo a concerted conformational transition, with 7 trajectories
showing multiple transitions.
Hence we have shown that the allosteric transition in PDZ3 proceeds in
a concerted manner. That is, all contacts change almost simultaneously, 
similar to the cooperative mechanism that was observed in
the functional dynamics of T4 lysozyme.\cite{Post22a} Starting in the
relatively ordered {\em cis} state, the system becomes disordered upon
{\em cis}-to-{\em trans} photoswitching and returns to an ordered
state via a cooperative transition. In this way, the system follows a
order-disorder-order transition, which was also observed for a
photoswitchable PDZ2 domain.\cite{Buchenberg17}

Comparing the distance distributions from the nonequilibrium and the
{\em trans} simulations (Fig.\ \ref{fig:contacts}a), we notice that
within $1\,\mu$s the system has not yet reached the {\em trans}
equilibrium state, because most contacts are less likely in {\em
  trans} than during the nonequilibrium process. Similarly as shown
for a photoswitchable PDZ2 domain,\cite{Buchenberg17} we thus find
that the final step of the allosteric transition is the slow
relaxation into the new equilibrium state. Interestingly, a visual
inspection of the experimental data (Fig.\ 3f in Ref.\
\onlinecite{Bozovic21}) also seems to indicate weak changes of the
transient infrared spectrum at long times $ t \gtrsim 1\,\mu$s.

%
%
\vspace*{-4mm}
\section*{Discussion and conclusions}
\vspace*{-4mm}

In their beautiful experiment, Bozovic et al.\cite{Bozovic21}
implemented an azobenzene photoswitch in the $\alpha_3$-helix of PDZ3,
which allows to change between the {\em cis} state ($\alpha_3$-helix
folded) and the {\em trans} state ($\alpha_3$-helix unfolded). The
underlying idea of the experiment is that the unfolding of $\alpha_3$
causes the separation of $\alpha_{3}$ from the protein core, which in
turn causes a change in ligand binding affinity,
\cite{Petit09,Bozovic20a} thus establishing an allosteric coupling
between $\alpha_3$ and the distal ligand. Using time-resolved infrared
spectroscopy, Bozovic et al.\ observed a timescale of 4 to 6\,ns
associated with the enforced unfolding of the $\alpha_3$-helix, as
well as a response of the ligand within 200\,ns.

Our simulation results reproduce these timescales quite accurately. We
obtain a timescale of 5\,ns associated with the initial stretching of
the $\alpha_{3}$-helix (Fig.\ \ref{fig:Ca}b), as well as a timescale
of $\sim 300\,$ns that reflects the reordering of the contact network
between the $\alpha_{3}$-helix and the core of the protein as well as
the ligand (Fig.\ \ref{fig:contacts}b).
Because the transition leads to a change of contacts formed by the
ligand, it eventually can be measured via vibrational spectroscopy of
the ligand. Hence we have shown that the measured timescale of 200\,ns
can be clearly assigned to elementary contact changes in the
protein. This also aids the interpretation of previous transient
infrared experiments on photoswitchable proteins,
\cite{Buchli13,Bozovic20} which too reported a typical timescale of
several hundreds of nanoseconds for this process.

To identify the intramolecular interactions mediating the allosteric
transition in PDZ3, we have employed a recently proposed correlation
analysis of interresidue contact distances.\cite{Diez22} It reveals a
network of highly correlated $\alpha_3$-protein contacts (Fig.\
\ref{fig:Leiden}b) that mediate a cooperative contact rearrangement in
PDZ3 (Fig.\ \ref{fig:SingleTraj}). Involving only a one-step
contact changing process, the cooperative transition can be considered
as a minimal solution to transduce the information on the state
of the $\alpha_3$-helix to the ligand, and represents therefore the
elementary step in the allosteric communication in PDZ3.
Cooperative contact rearrangements also cause large global
free energy barriers that offer a plausible explanation
why the often tiny structural changes in allostery can be stabilized
in spite of the presence of thermal fluctuations and noise.

Allosteric phenomena in PDZ domains (and in particular in PDZ3) have
been debated for long.\cite{Lockless99,
  Ota05,Kong09,Gerek11,Ishikura15, Kumawat17, Stock18,Faure22}
Nonetheless, our simulation of a specific experiment on
PDZ3 has revealed several new and in part unexpected findings.
To begin with, this concerns the existence of direct contacts between
the $\alpha_{3}$-helix and the ligand (Fig.\
\ref{fig:contacts}a). While the centers of mass of $\alpha_{3}$ and
the ligand are on average separated by about 1.5\,nm, the long
side-chains of Lys103, Azo102, Glu(3) and Lys(-4) nevertheless
facilitate such contacts. They are formed in particular during the
cooperative transition (Fig.\ \ref{fig:SingleTraj}), but may also
occur in the {\em trans} state.
Our results are in contrast to the NMR study of Petit et
al.,\cite{Petit09} which excluded the
possibility of direct $\alpha_3$-ligand contacts. However, they used a
different ligand (NYKQTSV instead of KETWV used here) and their PDZ3
was one amino acid shorter than ours (i.e, it misses Lys103 which is
involved in several contacts). 
We also note that direct $\alpha_3$-ligand contacts where discussed
for the slightly longer ligand KKETWV\cite{Murciano-Calles14} and for
a PDZ3 domains with a longer C-terminus. \cite{Chi12,Mostarda12}
Hence we conclude that the existence of direct contacts in PDZ3
are specific to the considered version of protein and ligand.

Recognizing that direct contacts between the putative active site (the
$\alpha_3$-helix) and the allosteric site (the binding pocket) may
cause at least in part the observed binding affinity change, one may
question if the notion of allostery is still adequate here. On the
other hand, we expect that the main finding, i.e., that allosteric
communication is mediated through a concerted conformational
transition that involves numerous simultaneous contact changes,
prevails also in the absence of direct contacts and therefore
represents a general result. For example, Post et al.\cite{Post22a}
found a quite similar scenario for the allosteric coupling in T4
lysozyme.

A further complication revealed by our analysis is that there is no
simple one-to-one relation of the {\em cis} and {\em trans} states of
PDZ3 with the attachment and the separation of $\alpha_{3}$,
respectively. Rather the calculations of the distribution of the
C$_\alpha$-distance $d_{29,102}$ (Fig.\ \ref{fig:Ca}a) and the
$\alpha_3$-ligand contact distances (Fig.\ \ref{fig:contacts}a) reveal
the coexistence of attached and separated structures in both states.
We obtain a population shift of 68 to 32\% of attached structures when
we change from {\em cis} to {\em trans}, while we find on average 51\%
during our nonequilibrium simulations (Fig.\ \figCaTimeEvol).
Interestingly, this dynamical heterogeneity of $\alpha_3$
occurs also in the wild-type system, where the attached structures
persist 45\% of the time (Fig.\ \figCaTimeEvol). Unfortunately,
existing NMR \cite{Petit09} or mutational \cite{Faure22} studies do
not include the $\alpha_{3}$-helix, which could confirm this effect.

Allosteric interactions have been commonly described by network
models, which may be based on statistical inference, \cite{Lockless99}
correlations, \cite{Sethi09} forces, \cite{Stacklies11} energy
transfer, \cite{Ishikura15} electrostatic interactions\cite{Kumawat17}
and many variants of thereof. While in this work we also have
constructed a correlation matrix reflecting allosteric interactions,
our work differs from common network approaches in two important
aspects. First, instead of calculating all interresidue correlations
using Cartesian C$_\alpha$-coordinates, we directly focus on
interresidue contacts that change during the allosteric
transition. For the considered conformational rearrangement in PDZ3,
the block diagonalization of the contact distances yielded a single
cluster containing highly correlated $\alpha_3$-protein contacts
(Fig.\ \ref{fig:Leiden}b), indicating that only a small part of the
protein is involved in the allosteric transition.

Second, rather than describing allostery via the interpretation of
some network model, we have directly simulated the allosteric
transition using nonequilibrium MD, and employed the correlation
analysis merely for the interpretation of the simulations. While such
an approach is desirable, we admit that it will be limited to rather
small allosteric systems. On the other hand, by a comparison to our MD
simulations, we are now in a position to test the validity and
potential of various network models, which may then facilitate the
study of larger proteins.


\vspace*{-6mm}
\subsection*{Acknowledgments}
\vspace*{-4mm} The authors thank Andrew L. Lee for providing
experimental order parameters, and Peter Hamm and his group as well as
Georg Diez, Emanuel Dorbath, Daniel Nagel and Matthias Post for
numerous instructive and helpful discussions. This work has been
supported by the Deutsche Forschungsgemeinschaft (DFG) via the
Research Unit FOR 5099 ''Reducing complexity of nonequilibrium''
(project No. 431945604). The authors acknowledge support by the High
Performance and Cloud Computing Group at the Zentrum f\"ur
Datenverarbeitung of the University of T\"ubingen and the
Rechenzentrum of the University of Freiburg, the state of
Baden-W\"urttemberg through bwHPC and the DFG through Grant Nos. INST
37/935-1 FUGG (RV bw16I016) and INST 39/963-1 FUGG (RV bw18A004).
\vspace*{-7mm}

\subsection*{Supporting Information Available:}
\vspace*{-4mm} Methods, 1 table, and  5 figures.

%
%

\end{document}


\setstretch{1}
	\title{Supplementary Information: Nonequilibrium Modeling of the Elementary Step in PDZ3 Allosteric Communication }
	\author{Ahmed A.\ A.\ I.\ Ali}
\altaffiliation{A.\ Ali and A.\ Gulzar contributed equally to this work.}
	\author{Adnan Gulzar}
\altaffiliation{A.\ Ali and A.\ Gulzar contributed equally to this work.}
	\author{Steffen Wolf}
	\author{Gerhard Stock}
\email{stock@physik.uni-freiburg.de}
\affiliation{Biomolecular Dynamics, Institute of Physics,
		University of Freiburg, 79104 Freiburg, Germany}
	\date{\today}
	\maketitle
	\baselineskip5mm
	
	\section*{Methods}
	\subsection*{Molecular dynamics simulations}
	
	All MD simulations of PDZ3 were performed using the GROMACS v2020 software package\cite{Abraham15} with a hybrid GPU-CPU acceleration scheme\cite{Pronk13}  and the Amber99*ILDN force field.\cite{Hornak06, Best09, LindorffLarsen10} Force field parameters of the azobenzene photoswitch were previously reported in  Ref.~\cite{Zanobini18}. Protein-ligand structures were solvated with ca. 6000 TIP3P water molecules\cite{Jorgensen83} in a dodecahedron box with a minimal image distance of 7 nm. 21 Na$\textsuperscript{+}$ and 19 Cl$\textsuperscript{-}$ were added to yield a charge-neutral system with a salt concentration of 0.15 M. All bonds involving hydrogen atoms were constrained using the LINCS algorithm\cite{hess2008p}, allowing for a time step of 2 fs. Long-range electrostatic interactions were computed by the Particle Mesh Ewald method\cite{Essmann95}, while the short-range electrostatic interactions were treated explicitly with the Verlet cutoff scheme. The minimum cutoff distance for electrostatic and van der Waals interactions was set to 1.4 nm. A temperature of 300 K was maintained via the Bussi thermostat\cite{Bussi07} (aka velocity-rescale algorithm) with a coupling time constant of $\tau_{T}$ = 0.1 ps. A pressure $P = $1 bar was controlled using the pressure coupling method of Berendsen\cite{Berendsen84} with a coupling time constant of $\tau_{P}$ = 0.1 ps.\\
The starting structure of the PDZ3 was taken from PDB entry 1TP5\cite{sarostructure}. To generate initial structures for the MD simulations of $\alpha_{3}$-switched PDZ3, we first removed the last 13 residues (104-116) from the crystal structure, and the azobenzene photoswitch was then attached at positions 95 and 102.\\ 
These residues were first mutated to cysteins as in the experiment of Bozovic et al.\cite{Olga21} to provide covalent connection points. Similarly, to be consistent with the experiments, we considered a 5-mer peptide (KETWV). The five amino acid peptide is the shortest possible sequence, which still includes the five crucial amino acids for specific binding to the PDZ3 domain.\cite{Saro07} \\
For the equilibrium simulations, PDZ3 was minimized in both $\textbf{\textit{cis}}$ and $\textbf{\textit{trans}}$ conformations of photoswitch. Following NPT equilibration of both systems for 10~ns, systems were again equilibrated in NVT-ensemble for 100~ns. From the last 50~ns, 8 statistically independent structures  (i.e., with different initial velocity distributions) were chosen for NVT runs of 5~$\rm{\mu}$s length each. In this way, we simulated both equilibrium systems for the cumulative time of 80~$\rm{\mu}$s each. For the nonequilibrium $\textbf{\textit{cis-to-trans}}$ simulations, we chose 100 statistically independent structures from last 50~ns of the NVT equilibration run of the $\textbf{\textit{cis}}$  state. Azobenzene photoswitching a performed by using potential-energy surface switching approach \cite{Nguyen06}. All 100 $\textbf{\textit{cis-to-trans}}$ nonequilibrium simulations were run for 1~$\mu$s.\\
Gromacs tools $gmx~mdmat$ and $gmx~mindist$ were employed to compute contact map, interresidue $C_{\alpha}$-distances, and the number of contacts between various segments of PDZ2. Time-dependent distributions and mean values of any observables were calculated via an ensemble average over 100 nonequilibrium trajectories. We consider a contact to be formed when the average minimum distance between two residues is less than 0.45 nm.\cite{Ernst15}
	
	\subsection*{Calculations of NMR order parameters}
	
	To calculate order parameters, we performed 10 additional NVT simulations for wild-type PDZ3, \textbf{cis} and \textbf{trans} conformations of photoswitchable PDZ3. Force filed Amber99SB*-ILDN with re-parameterization of methyl rotation barriers of Hoffmann\cite{hoffmann2018accurate} was used. MD simulations were performed for 200~ns, generating a cumulative time length of 2 $\mu$s for each system. In order to improve the prediction of calculated order parameters, frames of trajectories were printed after every 1 ps. Order parameters were calculated using the methods of Chen\cite{Chen18} and Hoffmann\cite{Hoffmann20}. The translational and rotational motions were removed  in order to calculate only the internal time correlation functions (TCFs) for each methyl group. 
	\begin{equation}
	\label{TCF_md}
	C(\tau) = \langle P_2[\vec{\mu}(0) \cdot \vec{\mu}(\tau)] \rangle
	\end{equation}
	$P_2$ is the second Legendre polynomial\\
	where
	\begin{equation}
	P_2(x) = \frac{1}{2}(3x^2-1) ,
	\end{equation}
	and $\vec{\mu}(\tau)$ is the unit bond vector (C--H bond vector in case of methyl group), and $\langle ... \rangle$ indicates averaging over all time step differences $(\tau)$.\\
	
	Finally, the TCFs were then fitted to LS2 (two parameters fit) model of Lipari-Szabo\cite{lipari1982model}:
	\begin{equation}
	\label{ls2}
	C_{LS2}(\tau) = S^2 + (1 - S^2) \, e^{-\frac{\tau}{\tau_f}} ,
	\end{equation}
	where $S^2 = S^2_f~S^2_{axis}$, with the order parameter of the symmetrical methyl rotation axis \cite{Kay91} $S^2_{axis}$ and $S^2_f$: 
	\begin{equation}
	\label{S_f2}
	S_f^2=P_2(\cos \, \theta_{ij}) \cdot P_2(\cos \, \theta_{kl}) 
	\end{equation}
	where $\theta_{ij}$ is the angle between two C--H bond vectors in the methyl group. For a methyl group with ideal tetrahedral geometry, $\theta_{ij} = 109.5^{\circ}$ and $P_2(\cos \, \theta_{ij}) = -1/3$,  we obtain $S^2_f = 1/9$.
	\subsection*{Contacts selection}
	To identify the contacts, we used $2.5 \times 10^5$ cis, $2.5 \times 10^5$ trans, and $5 \times 10^5$ NEQ frames to weight both equilibrium and nonequilibrium data equally. Moreover, we selected the contacts based on the following criteria:
	\begin{itemize}
		\item Minimal distance ($d_{ij}$) should be $\leq$ 0.45 nm between residues $i$'s and $j$'s non-hydrogen atoms.
		\item The two residues should be in contact for $\geq$ 10 \% of the considered frames.
		\item $|i-j| > 2 $ to neglect nearest and next to nearest neighbor residues.
	\end{itemize}
	These criteria resulted in 403 contact distances.
\clearpage
\section*{Results}	
	\begin{figure}[h!]
		\centering
		{\includegraphics[width=0.65\textwidth]{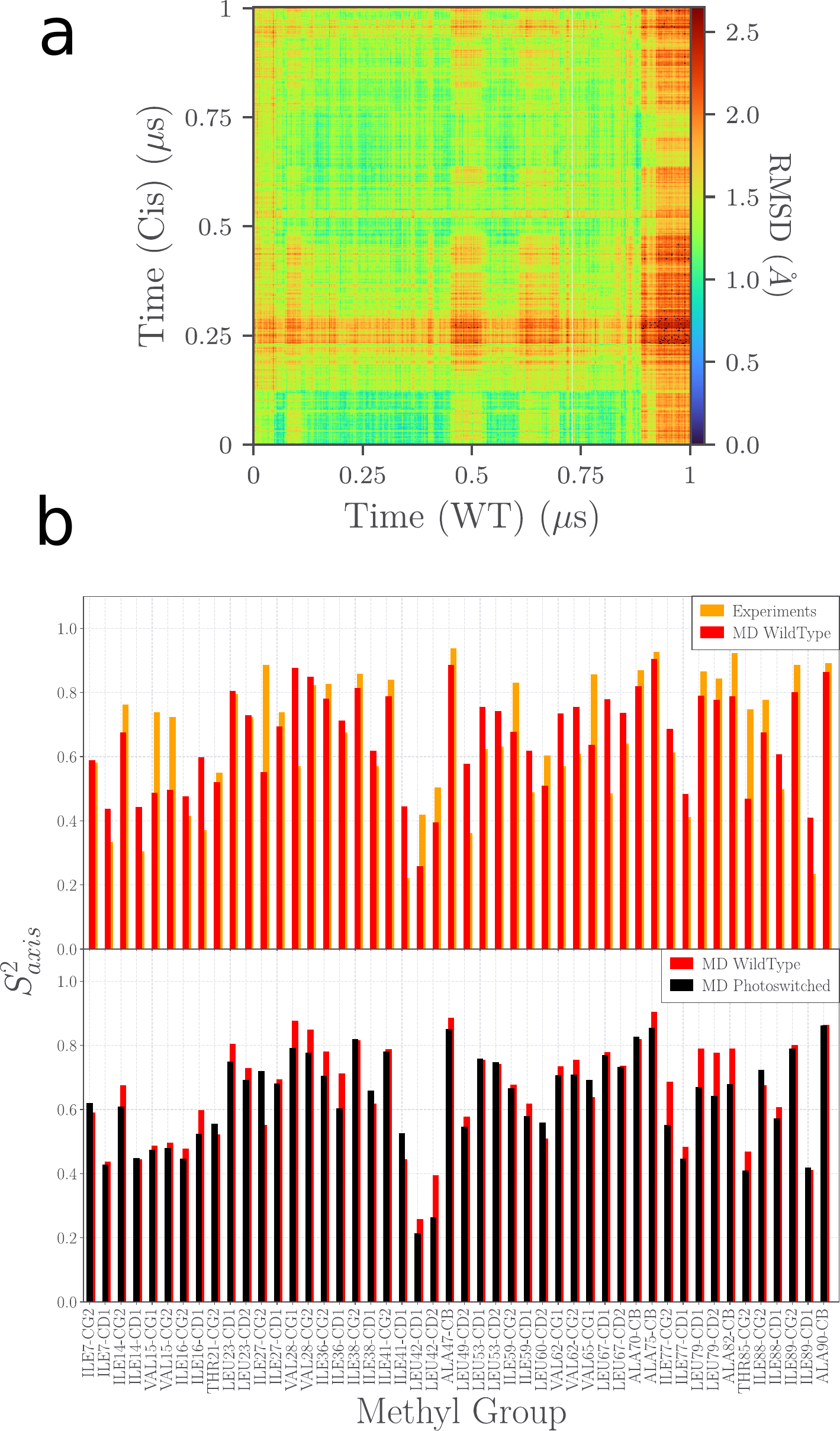}}
		\caption{(a) RMSD map between wild-type (WT) PDZ3 (x-axis) and
                  cis-PDZ3 (y-axis) for a 1 $\mu s$ equilibrium
                  trajectory. The RMSD was calculated for $C_\alpha$
                  atoms, excluding the ligand and the first 5 and the
                  last 2 residues of the protein, because of their
                  high flexibility. The reference structure was set as
                  the first frame of the WT simulation. This matrix
                  has an average RMSD value of 1.45 $\AA$. (b) Bar
                  chart between (top) experimental \cite{petit09} and
                  computational (WT PDZ3) results for
                  $S^2_{axis}$. (bottom) Results for $S^2_{axis}$ WT
                  and photoswitched PDZ3 (cis) calculations.}
		\label{figRMSD}
	\end{figure}

\begin{figure}[h!]
\centering
{\includegraphics[width=0.8\textwidth]{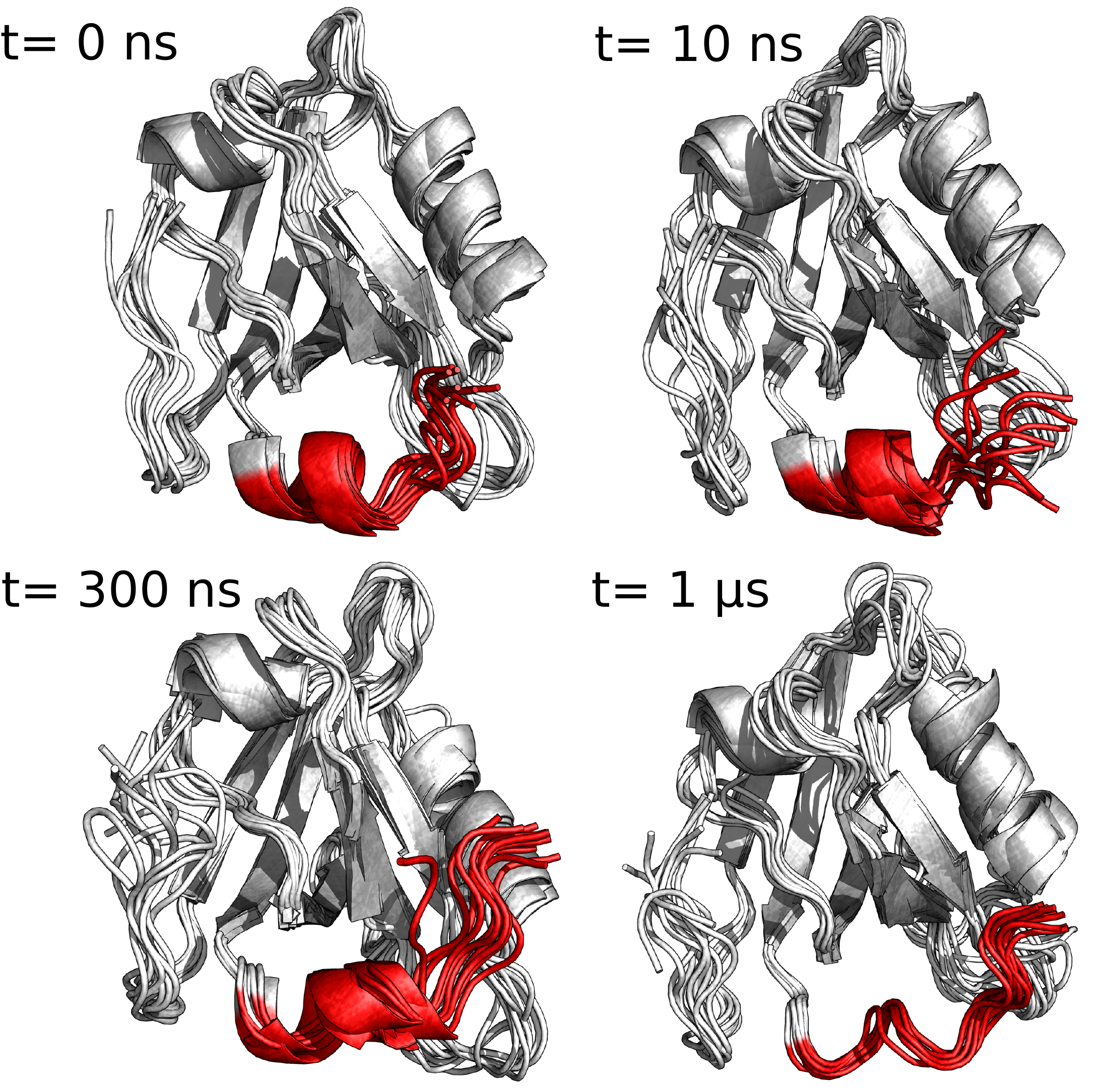}}
\caption{Enforced unfolding of the $\alpha_3$-helix (red) of PDZ3,
  following cis-to-trans photoswitching of $\alpha_3$ (see Fig.\ 1 of
  the main text). Shown are snapshots of 10 randomly chosen 
nonequilibrium MD trajectories at indicated simulation times.}
\end{figure}

\begin{figure}[h!]
	\centering
	{\includegraphics[width=1\textwidth]{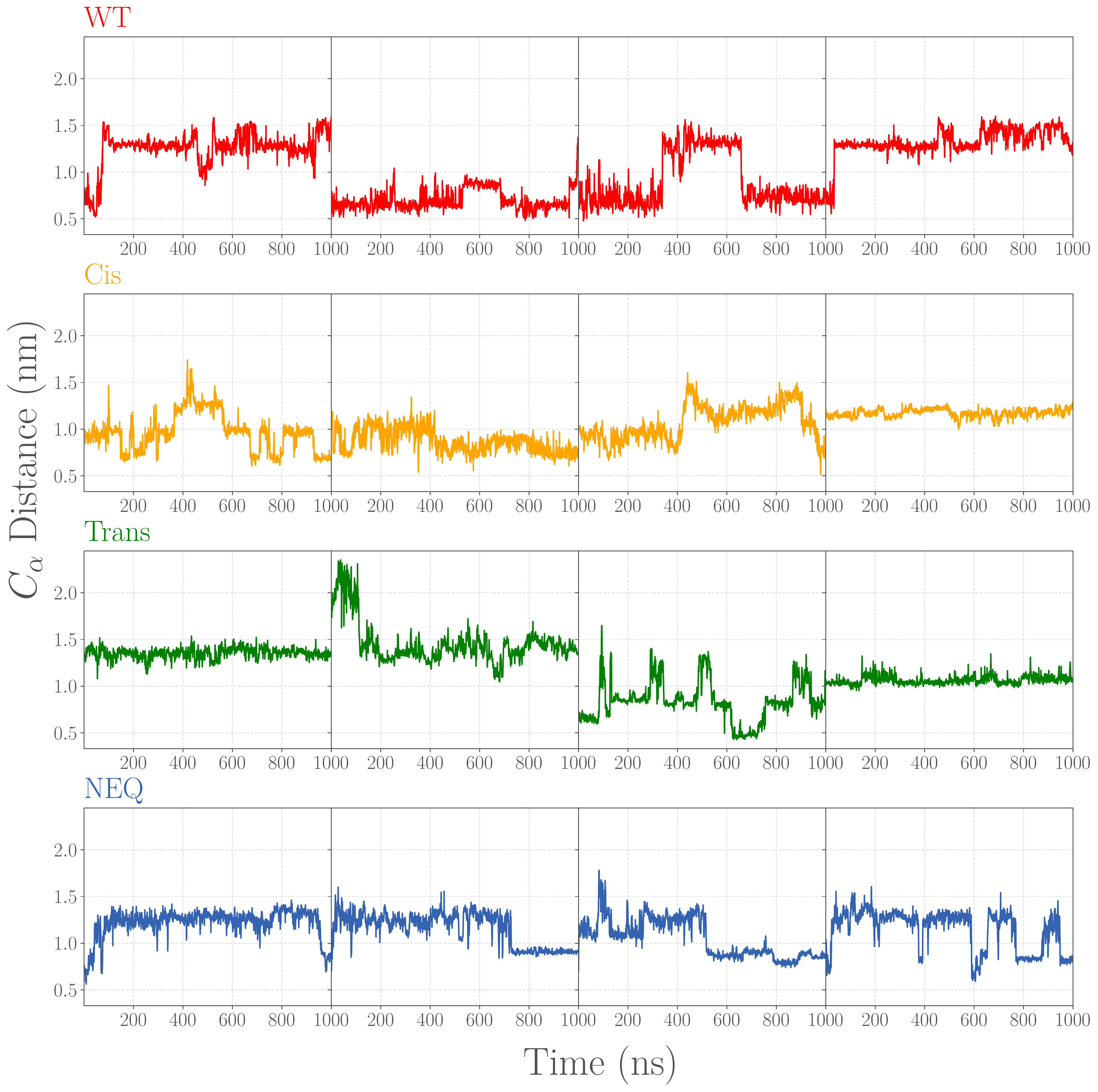}}
	\caption{Time evolution of $C_{\alpha}$ $d_{29, 102}$ distance
          between $\alpha_3$ and $\beta_2\beta_3$ obtained for
          wild-type, cis, trans and nonequilibrium systems. For
          smoothing a Gaussian filter with width $\sigma = 0.2\,$ns
          was used.}
	\label{29, 102}
\end{figure}

\renewcommand*{\arraystretch}{1.5}
\begin{table}[bh!]
	\label{table:res_5}
	\caption{Coordinates of Leiden clusters 1 - 8 shown in Fig.\ 3
          of the main text. Clusters with a star indicate coordinates
          (or mini-clusters with $\le 3$ coordinates) that are not a
          member of a main cluster but still exhibit moderate
          correlation with some of the main clusters.}
	\begin{tabular}{ll}
		\toprule
		Cluster & Coordinates \\
		\midrule 
		1& $r_{102, (-3)}$, $r_{28, 102}$, $r_{39, 102}$, $r_{28, 101}$, $r_{97, 102}$, $r_{101, (-4)}$, $r_{101, (-3)}$, $r_{94, 102}$, $r_{103, (-4)}$, $r_{55, 102}$,\\& $r_{103, (-3)}$\\
		1$^*$&$r_{32, 70}$, $r_{32, 71}$, $r_{32, 69}$, $r_{55, 97}$, $r_{98, 102}$, $r_{5, 95}$, $r_{29, 34}$, $r_{94, 98}$, $r_{95, 99}$, $r_{93, 97}$, $r_{39, 97}$\\
		2& $r_{29, 100}$, $r_{37, 100}$, $r_{35, 100}$, $r_{30, 100}$, $r_{28, 100}$, $r_{97, 100}$, $r_{34, 100}$, $r_{31, 100}$, $r_{96, 100}$\\
		2$^*$& $r_{97, 101}$, $r_{96, 99}$, $r_{100, 103}$, $r_{99, 102}$\\
		3&$r_{35, 69}$, $r_{35, 68}$, $r_{33, 69}$, $r_{34, 69}$, $r_{35, 67}$, $r_{33, 70}$, $r_{34, 68}$, $r_{58, 68}$, $r_{33, 71}$, $r_{35, 70}$, $r_{59, 68}$, $r_{33, 68}$,\\& $r_{29, 71}$, $r_{36, 79}$\\
		3$^*$&$r_{34, 70}$, $r_{60, 68}$, $r_{61, 68}$, $r_{60, 67}$, $r_{28, 72}$, $r_{27, 72}$, $r_{67, 78}$, $r_{62, 86}$, $r_{36, 75}$, $r_{59, 79}$, $r_{73, (-2)}$, $r_{76, (-1)}$,\\& $r_{30, 72}$, $r_{36, 71}$, $r_{36, 70}$, $r_{79, 86}$, $r_{70, 74}$, $r_{69, 74}$, $r_{34, 96}$, $r_{66, 70}$, $r_{28, 38}$, $r_{42, (-1)}$, $r_{30, 34}$, $r_{28, 35}$,\\& $r_{31, 73}$, $r_{30, 36}$, $r_{30, 70}$, $r_{30, 35}$, $r_{67, 79}$\\
		4&$r_{28, (-4)}$, $r_{27, (-3)}$, $r_{29, (-4)}$, $r_{27, (-4)}$, $r_{28, (-3)}$, $r_{72, (-4)}$\\
		4$^*$&$r_{72, (-2)}$, $r_{27, (-2)}$, $r_{76, (-2)}$, $r_{72, (-3)}$, $r_{31, (-4)}$, $r_{39, (-3)}$\\
		5& $r_{78, 82}$, $r_{62, 82}$, $r_{82, 86}$, $r_{79, 82}$, $r_{63, 82}$, $r_{81, 86}$, $r_{77, 81}$, $r_{79, 83}$, $r_{18, 81}$, $r_{83, 86}$\\
		5$^*$& $r_{63, 81}$, $r_{82, 85}$, $r_{80, 83}$, $r_{62, 78}$, $r_{63, 78}$, $r_{59, 62}$\\
		6& $r_{21, 25}$, $r_{20, (-1)}$, $r_{18, (0)}$, $r_{21, 43}$, $r_{21, (-1)}$, $r_{21, 45}$, $r_{21, 24}$, $r_{24, (0)}$, $r_{22, (-1)}$, $r_{18, (-1)}$, $r_{21, 42}$,\\& $r_{18, 82}$, $r_{21, 46}$, $r_{18, 83}$, $r_{18, 86}$\\
		6$^*$& $r_{25, 79}$, $r_{80, (0)}$, $r_{25, (0)}$, $r_{24, 46}$, $r_{24, 45}$, $r_{26, 42}$, $r_{25, 42}$, $r_{24, (-1)}$, $r_{27, (0)}$, $r_{17, 85}$, $r_{25, (-2)}$, $r_{17, 22}$,\\& $r_{23, (0)}$, $r_{23, 45}$, $r_{25, 47}$, $r_{24, 40}$, $r_{19, 22}$, $r_{18, 21}$, $r_{19, 84}$, $r_{19, 46}$, $r_{70, 75}$\\
		7& $r_{47, 52}$, $r_{47, 51}$, $r_{16, 52}$, $r_{25, 53}$, $r_{15, 52}$, $r_{14, 50}$, $r_{50, 53}$, $r_{51, 54}$\\
		7$^*$&  $r_{23, 47}$, $r_{12, 54}$, $r_{47, 50}$, $r_{52, 57}$, $r_{41, 48}$, $r_{12, 91}$\\
	 	8& $r_{1, 95}$, $r_{2, 95}$, $r_{1, 98}$, $r_{2, 51}$, $r_{1, 51}$, $r_{1, 102}$, $r_{2, 98}$, $r_{2, 52}$, $r_{3, 54}$, $r_{2, 102}$, $r_{1, 100}$, $r_{1, 48}$, $r_{3, 95}$, $r_{2, 94}$, $r_{2, 53}$,\\& $r_{2, 54}$, $r_{2, 97}$\\
	 	8$^*$& $r_{4, 55}$, $r_{3, 55}$, $r_{3, 97}$, $r_{2, 55}$\\
		\bottomrule
	\end{tabular}
\end{table}
\begin{table}[bh!]
\begin{tabular}{ll}
	\toprule
	Cluster & Coordinates \\
	\midrule
	Noise& $r_{29, 35}$, $r_{34, 58}$, $r_{3, 92}$, $r_{31, 72}$, $r_{31, 71}$, $r_{30, 71}$, $r_{29, 72}$, $r_{25, 59}$, $r_{27, 59}$, $r_{76, (0)}$, $r_{36, 72}$, $r_{27, 75}$, $r_{27, 76}$,\\& $r_{36, 67}$, $r_{79, 88}$, $r_{62, 79}$, $r_{71, 74}$, $r_{31, 34}$, $r_{31, 70}$, $r_{30, 33}$, $r_{28, 36}$, $r_{29, 36}$, $r_{35, 58}$, $r_{36, 57}$, $r_{36, 58}$, $r_{36, 59}$,\\& $r_{73, (-4)}$, $r_{40, (-3)}$, $r_{40, (-1)}$, $r_{26, (-3)}$, $r_{26, (-1)}$, $r_{78, 81}$, $r_{23, 79}$, $r_{27, 79}$, $r_{67, 74}$, $r_{74, 77}$, $r_{71, 75}$, $r_{73, 77}$,\\& $r_{73, 76}$, $r_{65, 78}$, $r_{67, 75}$, $r_{62, 67}$, $r_{72, 76}$, $r_{72, 75}$, $r_{80, (-1)}$, $r_{80, (-2)}$, $r_{76, 80}$, $r_{76, 79}$, $r_{77, 80}$, $r_{79, (0)}$,\\& $r_{75, 78}$, $r_{75, 79}$, $r_{74, 78}$, $r_{59, 67}$, $r_{61, 67}$, $r_{66, 69}$, $r_{61, 89}$, $r_{67, 70}$, $r_{26, 37}$, $r_{27, 38}$, $r_{27, 39}$, $r_{27, 37}$, $r_{26, 38}$,\\& $r_{26, 39}$, $r_{28, 37}$, $r_{28, 39}$, $r_{24, 47}$, $r_{26, (-2)}$, $r_{25, (-1)}$, $r_{17, 46}$, $r_{17, 23}$, $r_{16, 25}$, $r_{18, 22}$, $r_{22, (0)}$, $r_{21, (0)}$,\\& $r_{23, 46}$, $r_{18, 46}$, $r_{22, 46}$, $r_{16, 23}$, $r_{18, 79}$, $r_{18, 80}$, $r_{18, 23}$, $r_{25, 40}$, $r_{26, 40}$, $r_{24, 41}$, $r_{24, 42}$, $r_{42, 45}$, $r_{45, 48}$,\\& $r_{16, 50}$, $r_{46, 50}$, $r_{45, 49}$, $r_{46, 49}$, $r_{44, 49}$, $r_{48, 51}$, $r_{61, 65}$, $r_{60, 65}$, $r_{61, 64}$, $r_{60, 66}$, $r_{61, 66}$, $r_{62, 65}$, $r_{63, 86}$,\\& $r_{63, 85}$, $r_{63, 87}$, $r_{61, 87}$, $r_{62, 87}$, $r_{4, 92}$, $r_{4, 54}$, $r_{1, 4}$, $r_{2, 5}$, $r_{5, 54}$, $r_{9, 54}$, $r_{54, 92}$, $r_{5, 9}$, $r_{12, 53}$, $r_{41, 47}$,\\& $r_{12, 52}$, $r_{53, 57}$, $r_{54, 57}$, $r_{57, 90}$, $r_{47, 53}$, $r_{53, 90}$, $r_{57, 91}$, $r_{57, 92}$, $r_{12, 92}$, $r_{56, 92}$, $r_{57, 93}$, $r_{56, 94}$, $r_{56, 93}$,\\& $r_{56, 91}$, $r_{55, 92}$, $r_{41, 55}$, $r_{41, 54}$, $r_{41, 53}$, $r_{25, 41}$, $r_{38, 41}$, $r_{38, 55}$, $r_{38, 56}$, $r_{38, 97}$, $r_{39, 55}$, $r_{40, 55}$, $r_{37, 57}$,\\& $r_{38, 57}$, $r_{38, 54}$, $r_{95, 98}$, $r_{28, 97}$, $r_{56, 97}$, $r_{34, 99}$, $r_{37, 99}$, $r_{37, 93}$, $r_{37, 56}$, $r_{34, 37}$, $r_{37, 97}$, $r_{94, 97}$, $r_{37, 58}$,\\& $r_{58, 96}$, $r_{93, 96}$, $r_{37, 96}$, $r_{8, 94}$, $r_{7, 94}$, $r_{6, 95}$, $r_{8, 95}$, $r_{5, 94}$, $r_{5, 92}$, $r_{7, 92}$, $r_{6, 94}$, $r_{8, 93}$, $r_{8, 92}$, $r_{9, 92}$, $r_{14, 90}$,\\& $r_{14, 89}$, $r_{12, 88}$, $r_{13, 52}$, $r_{14, 52}$, $r_{14, 53}$, $r_{13, 87}$, $r_{13, 88}$, $r_{14, 87}$, $r_{14, 88}$, $r_{23, 88}$, $r_{25, 88}$, $r_{61, 88}$, $r_{62, 88}$,\\& $r_{53, 88}$, $r_{59, 88}$, $r_{16, 88}$, $r_{16, 46}$, $r_{16, 47}$, $r_{23, 86}$, $r_{16, 53}$, $r_{11, 60}$, $r_{11, 89}$, $r_{11, 90}$, $r_{12, 89}$, $r_{12, 90}$, $r_{10, 90}$,\\& $r_{10, 91}$, $r_{9, 91}$, $r_{11, 91}$, $r_{58, 93}$, $r_{60, 91}$, $r_{58, 89}$, $r_{59, 90}$, $r_{60, 90}$, $r_{58, 90}$, $r_{59, 89}$, $r_{60, 89}$, $r_{58, 91}$, $r_{25, 38}$,\\& $r_{38, 53}$, $r_{27, 36}$, $r_{38, 59}$, $r_{44, 48}$, $r_{43, 48}$, $r_{42, 48}$, $r_{14, 86}$, $r_{15, 86}$, $r_{15, 87}$, $r_{15, 85}$, $r_{16, 85}$, $r_{16, 86}$, $r_{17, 84}$,\\& $r_{18, 84}$, $r_{16, 84}$, $r_{12, 57}$, $r_{13, 89}$, $r_{95, 102}$\\
	\bottomrule
	\end{tabular}
\end{table}
\renewcommand*{\arraystretch}{1}

	\begin{figure}
		\centering
		{\includegraphics[width=1\textwidth]{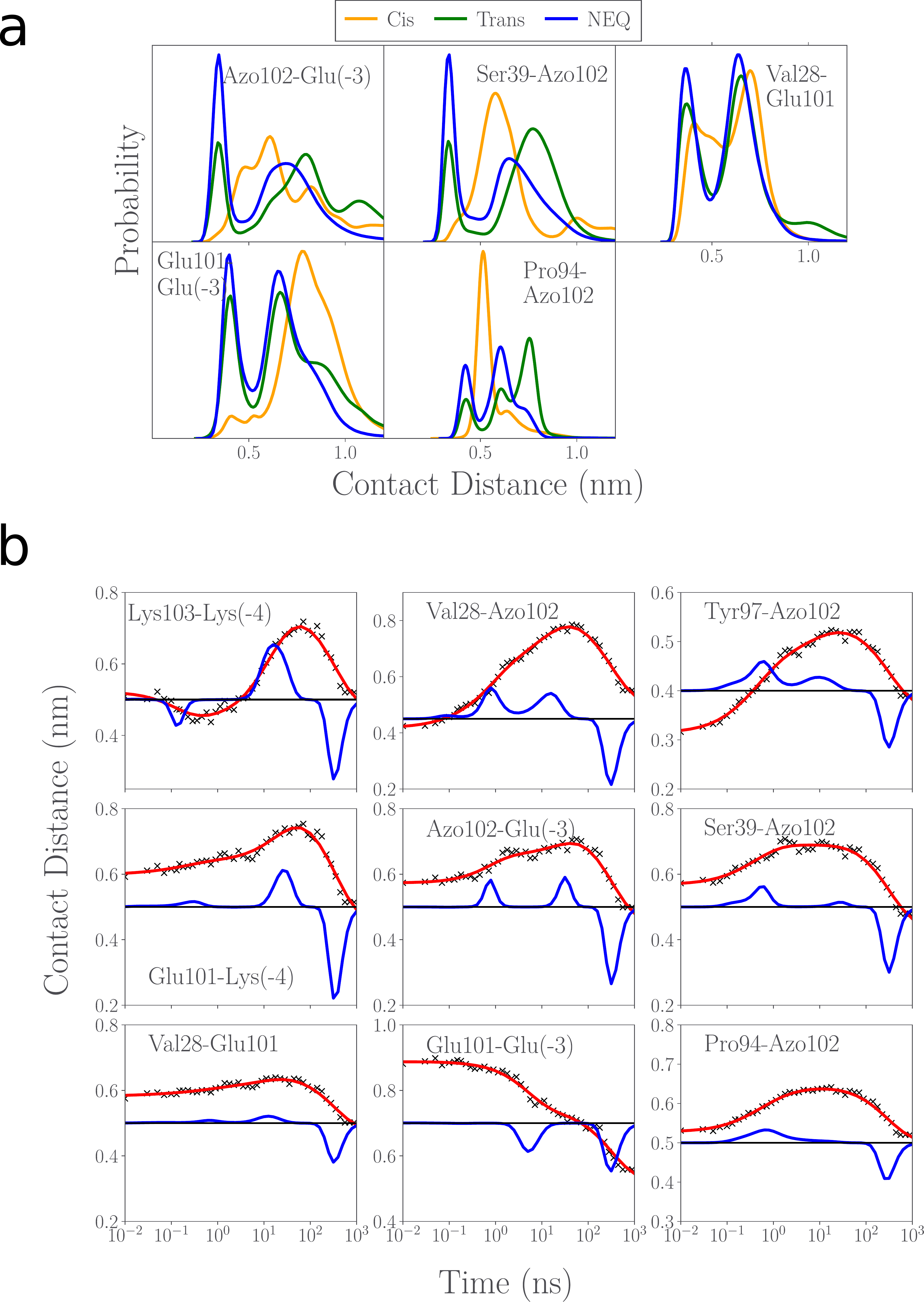}}
		\caption{(a) Probability distribution of 5 contact distances from cluster 1. (b) Timescale analysis of 9 contact distances from cluster 1, using noneqilibrium data.}
		\label{hist_timescale}
	\end{figure}

		
\begin{figure}
	\centering
	\caption{Time evolution of selected contacts that undergoes a
          nonequilibrium evolution from $cis$ to $trans$. For
          smoothing a Gaussian filter with width $\sigma = 0.2\,$ns
          was used.}
	{\includegraphics[scale=0.5]{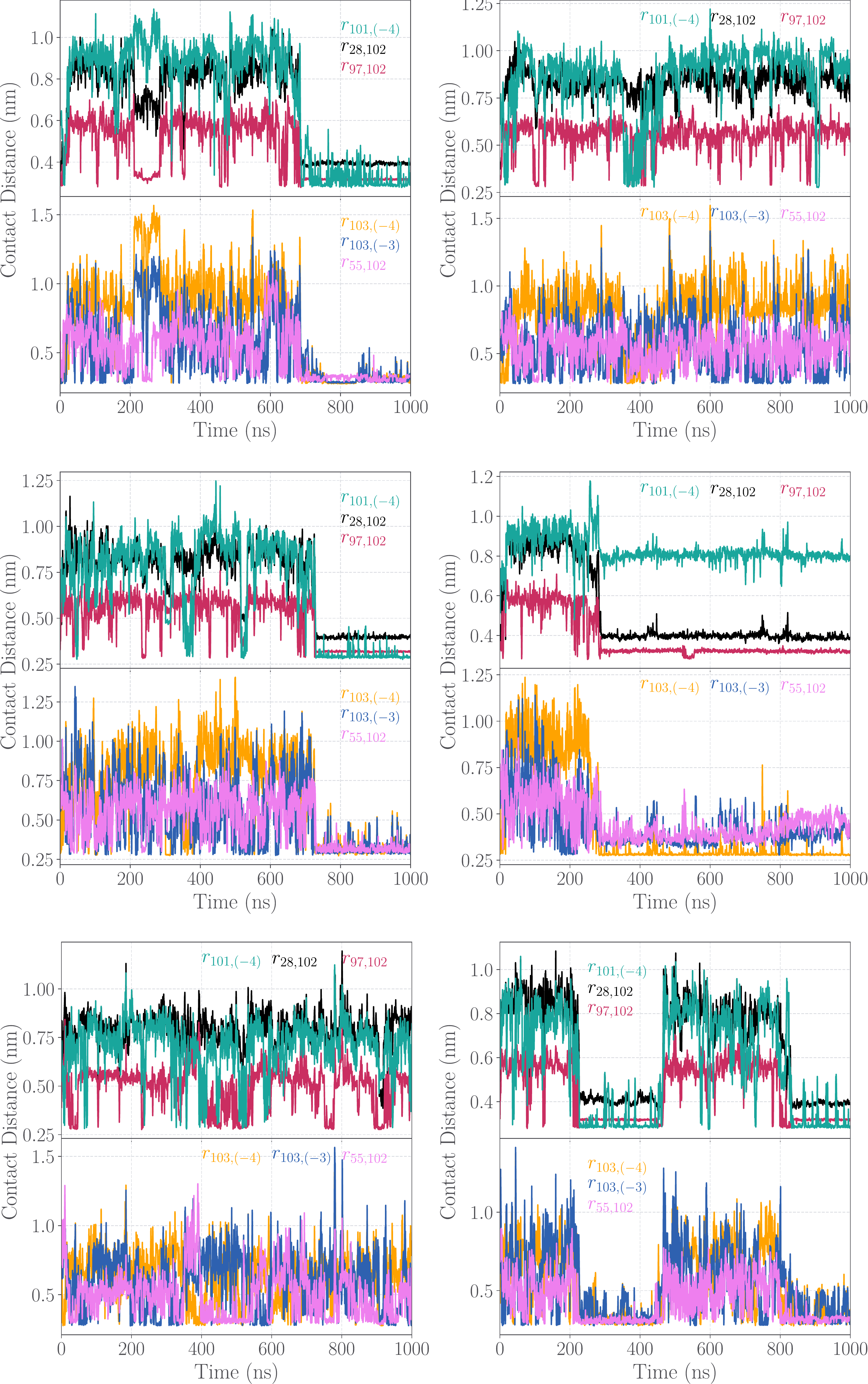}}
	\label{fig5_p1}
\end{figure}

\begin{figure}
	\centering
	{\includegraphics[scale=0.5]{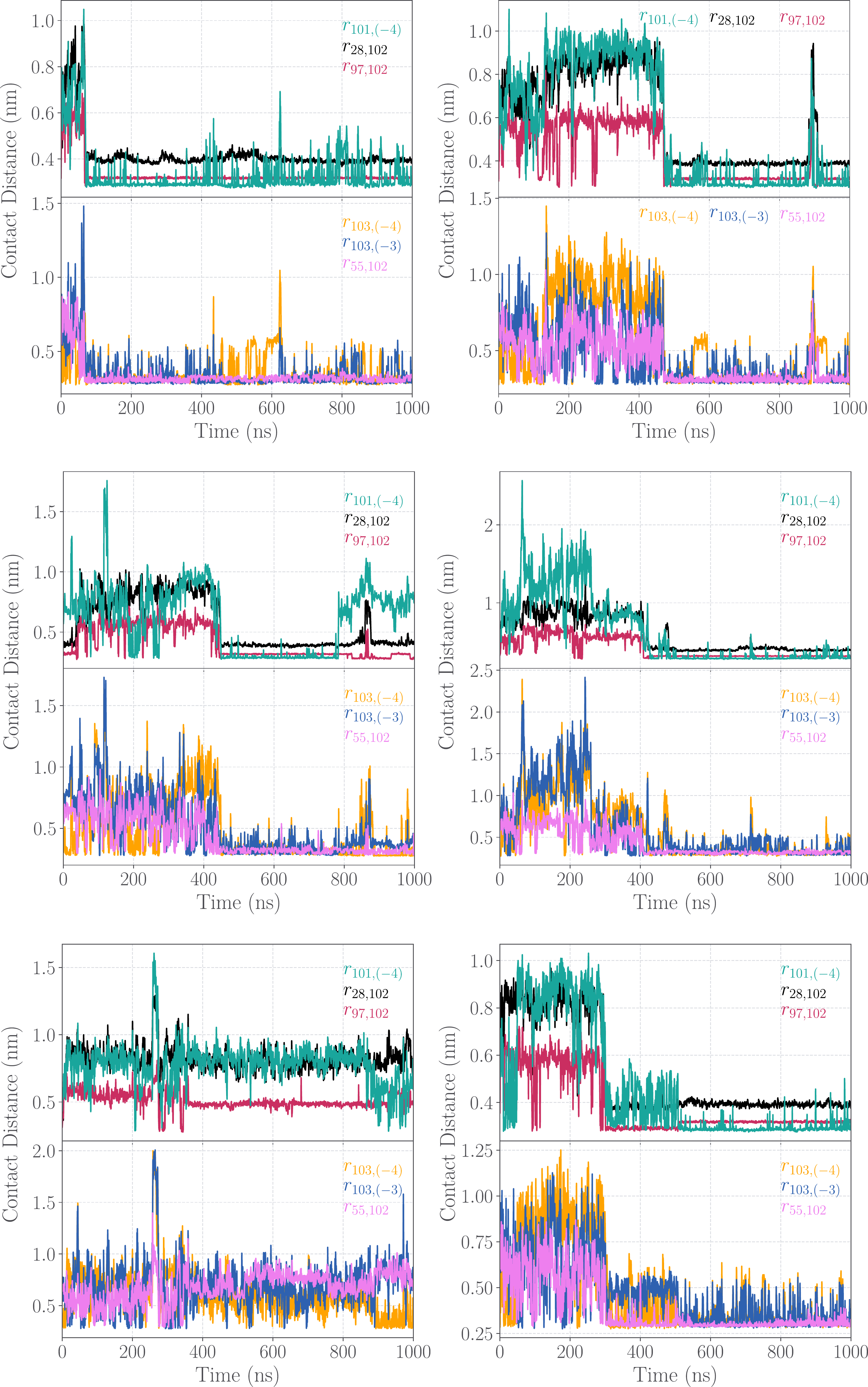}}
	\label{fig5_p2}
\end{figure}